# Low-Dose CT with a Residual Encoder-Decoder Convolutional Neural Network (RED-CNN)

Hu Chen, Yi Zhang*, *Member, IEEE*, Mannudeep K. Kalra, Feng Lin, Yang Chen, Peixo Liao, Jiliu Zhou, *Senior Member, IEEE*, and Ge Wang, *Fellow, IEEE*

*Abstract*—Given the potential risk of X-ray radiation to the patient, low-dose CT has attracted a considerable interest in the medical imaging field. Currently, the main stream low-dose CT methods include vendor-specific sinogram domain filtration and iterative reconstruction algorithms, but they need to access raw data whose formats are not transparent to most users. Due to the difficulty of modeling the statistical characteristics in the image domain, the existing methods for directly processing reconstructed images cannot eliminate image noise very well while keeping structural details. Inspired by the idea of deep learning, here we combine the autoencoder, deconvolution network, and shortcut connections into the residual encoder-decoder convolutional neural network (RED-CNN) for low-dose CT imaging. After patch-based training, the proposed RED-CNN achieves a competitive performance relative to the-state-of-art methods in both simulated and clinical cases. Especially, our method has been favorably evaluated in terms of noise suppression, structural preservation, and lesion detection.

## I. INTRODUCTION

X-RAY computed tomography (CT) has been widely utilized in clinical, industrial and other applications. Due to the increasing use of medical CT, concerns have been expressed on the overall radiation dose to a patient. The research interest has been strong in CT dose reduction under the well-known guiding principle of ALARA (as low as reasonably achievable) [1]. The most common way to lower the radiation dose is to reduce the X-ray flux by decreasing the operating current and shortening the exposure time of an X-ray tube. In general, the weaker the X-ray flux, the noisier a reconstructed CT image, which degrades the signal-to-noise ratio and could compromise the diagnostic performance. To address this inherent physical problem, many algorithms were designed to improve the image quality for low-dose CT (LDCT). These algorithms can be generally categorized into three categories: (a) sinogram domain filtration, (2) iterative reconstruction, and (3) image processing.

Sinogram filtering techniques perform on either raw data or log-transformed data before image reconstruction, such as filtered backprojection (FBP). The main convenience in the data domain is that the noise characteristic has been well known. Typical methods include structural adaptive filtering [2], bilateral filtering [3], and penalized weighted least-squares (PWLS) algorithms [4]. However, the sinogram filtering methods often suffer from spatial resolution loss when edges in the sinogram domain are not well preserved.

Over the past decade, iterative reconstruction (IR) algorithms have attracted much attention especially in the field of LDCT. This approach combines the statistical properties of data in the sinogram domain, prior information in the image domain, and even parameters of the imaging system into one unified objective function. With compressive sensing (CS) [5], several image priors were formulated as sparse transforms to deal with the low-dose, few-view, limited-angle and interior CT issues, such as total variation (TV) and its variants [6-9], nonlocal means (NLM) [10-12], dictionary learning [13], low-rank [14], and other techniques. Model based iterative reconstruction (MBIR) takes into account the physical acquisition processes and has been implemented on some current CT scanners [15]. Although IR methods obtained exciting results, there are two weaknesses. First, on most of modern MDCT scanners, IR techniques have replaced FBP based image reconstruction techniques for radiation dose reduction. However, these IR techniques are vendor-specific since the details of the scanner geometry and correction steps are not available to users and other vendors. Second, there are substantial computational overhead costs associated with popular IR techniques. Fully model-based iterative reconstruction techniques have greater

This work was supported in part by the National Natural Science Foundation of China under Grants 61671312, 61302028, 61202160, 81370040 and 81530060 and in part by the National Institute of Biomedical Imaging and Bioengineering/National Institutes of Health under Grants R01 EB016977 and U01 EB017140. Asterisk indicates corresponding author.

H. Chen, Y. Zhang*, F. Lin, and J. Zhou are with the College of Computer Science, Sichuan University, Chengdu 610065, China (e-mail: huchen@scu.edu.cn; yzhang@scu.edu.cn; linfeng@scu.edu.cn; zhoujl@scu.edu.cn).

M. K. Kalra is with Department of Radiology, Massachusetts General Hospital, Boston, MA 02114, USA (e-mail: mkalra@mgh.harvard.edu).

Y. Chen is with the Laboratory of Image Science and Technology, Southeast University, Nanjing 210096, China, and also with the Key Laboratory of Computer Network and Information Integration (Southeast University), Ministry of Education, Nanjing 210096, China (e-mail: chenyang.list@seu.edu.cn).

P. Liao is with Department of Scientific Research and Education, The Sixth People's Hospital of Chengdu, Chengdu 610065, China (e-mail: universe6527@163.com).

G.Wang is with Department of Biomedical Engineering, Rensselaer Polytechnic Institute, Troy, NY 12180 USA (e-mail: wangg6@rpi.edu).

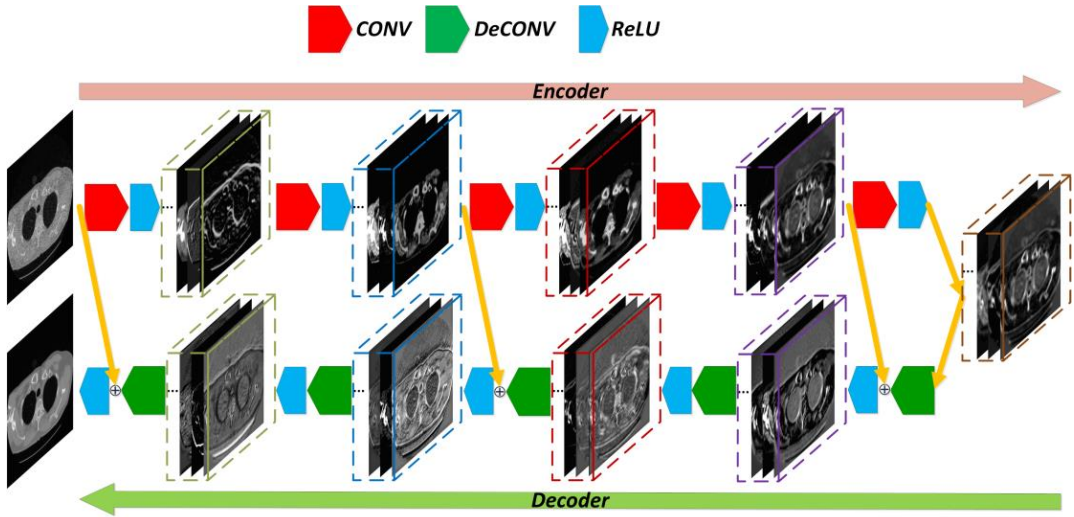

Fig.1. Overall architecture of our proposed RED-CNN network.

potential for radiation dose reduction but slow reconstruction speed and changes in image appearance limit their clinical applications.

An alternative for LDCT is post-processing of reconstructed images, which does not rely on raw data. These techniques can be directly applied on LDCT images, and integrated into any CT system. In [16], NLM was introduced to take advantage of the feature similarity within a large neighborhood in a reconstructed image. Inspired by the theory of sparse representation, dictionary learning [17] was adapted for LDCT denoising, and resulted in substantially improved quality abdomen images [18]. Meanwhile, block-matching 3D (BM3D) was proved efficient for various X-ray imaging tasks [19-21]. In contrast to the other two kinds of methods, the noise distribution in the image domain cannot be accurately determined, which prevents users from achieving the optimal tradeoff between structure preservation and noise supersession.

Recently, deep learning (DL) has generated an overwhelming enthusiasm in several imaging applications, ranging from low-level to high-level tasks from image denoising, deblurring and super resolution to segmentation, detection and recognition [22]. It simulates the information processing procedure by human, and can efficiently learn high-level features from pixel data through a hierarchical network framework [23].

Several DL algorithms have been proposed for image restoration using different network models [24-31]. As the autoencoder (AE) has a great potential for image denoising, stacked sparse denoising autoencoder (SSDA) and its variant were introduced [24-26]. Convolutional neural networks are powerful tools for feature extraction and were applied for image denoising, deblurring and super resolution [27-29]. Burger et al. [30] analyzed the performance of multi-layer perception (MLP) as applied to image patches and obtained competitive results as compared to the state-of-the-art methods. Previous studies also applied DL for medical image analysis, such as tissue segmentation [32, 33], organ classification [34] and nuclei detection [35]. Furthermore, reports started emerging on tomographic imaging topics. For example, Wang et al. incorporated a DL-based regularization term into a fast MRI reconstruction framework [36]. Chen et al. presented preliminary results with a light-weight CNN-based framework for LDCT imaging [37]. A deeper version using the wavelet transform as inputs was presented [38] which won the second place in the "*2016 NIH-AAPM-Mayo Clinic Low Dose CT Grand Challenge*." The filtered back-projection (FBP) workflow was mapped to a deep CNN architecture, reducing the reconstruction error by a factor of two in the case of limited-angle tomography [39]. An overall perspective was also published on deep learning, or machine learning in general, for tomographic reconstruction [40].

Despite the interesting results on CNN for LDCT, the potential of the deep CNN has not been fully realized. Although some studies involved construction of deeper networks [41, 42], most image denoising models had limited layers (usually 2~3 layers) since image denoising is considered as a "low-level" task without intention to extract features. This is in clear contrast to high-level tasks such as recognition or detection, in which pooling and other operations are widely used to bypass image details and capture topological structures.

Inspired by the work of [31], we incorporated a deconvolution network [43] and shortcut connections [41, 42] into a CNN model, which is referred to as a residual encoder-decoder convolutional neural network (RED-CNN). In the second section, the proposed network architecture is described. In the third section, the proposed model is evaluated and validated. In the final section, the conclusion is drawn.

## II. METHODS

### A. Noise Reduction Model

Our workflow starts with a straightforward FBP reconstruction from a low-dose scan, and the image denoising problem is restricted within the image domain [37]. Since the DL-based methods are independent of the statistical distribution of image noise, the LDCT problem can be simplified to the following one. Assuming that $\mathbf{X} \in \mathbf{R}^{m \times n}$ is a LDCT image and $\mathbf{Y} \in \mathbf{R}^{m \times n}$ is a corresponding normal dose CT (NDCT)

image, the relationship between them can be formulated as
$$\mathbf{X} = \sigma(\mathbf{Y}) \quad (1)$$
where $\sigma:\mathbf{R}^{m\times n} \to \mathbf{R}^{m\times n}$ denotes the complex degradation process involving quantum noise and other factors. Then, the problem can be transformed to seek a function $f$:
$$\arg\min_{f} \| f(\mathbf{X}) - \mathbf{Y} \|_2^2 \quad (2)$$
where $f$ is regarded as the optimal approximation of $\sigma^{-1}$, and can be estimated using DL techniques.

*B. Residual Autoencoder Network*

The autoencoder (AE) was originally developed for unsupervised feature learning from noisy inputs, which is also suitable for image restoration. In the context of image denoising, CNN also demonstrated an excellent performance. However, due to its multiple down-sampling operations, some image details can be missed by CNN. For LDCT, here we propose a residual network combining AE and CNN, which has an origin in the work [31]. Rather than adopting fully-connected layers for encoding and decoding, we use both convolutional and deconvolutional layers in symmetry. Furthermore, different from the typical encoder-decoder structure, residual learning [41] with shortcuts is included to facilitate the operations of the convolutional and corresponding deconvolutional layers. There are two modifications to the network described in [31]: (a) the ReLU layers before summation with residuals have been removed to abandon the positivity constraint on learned residuals; and (b) shortcuts have been added to improve the learning process.

The overall architecture of the proposed RED-CNN network is shown in Fig. 1. This network consists of 10 layers, including 5 convolutional and 5 deconvolutional layers symmetrically arranged. Shortcuts connect matching convolutional and deconvolutional layers. Each layer is followed by its rectified linear units (ReLU) [44]. The details about the network are described as follows.

*1) Patch extraction*

DL-based methods need a huge number of samples. This requirement cannot be easily met in practice, especially for clinical imaging. In this study, we propose to use overlapped patches in CT images. This strategy has been found to be effective and efficient, because the perceptual differences of local regions can be detected, and the number of samples are significantly boosted [24, 27, 28]. In our experiments, we extracted patches from LDCT and corresponding NDCT images with a fixed size.

*2) Stacked encoders (Noise and artifact reduction)*

Unlike the traditional stacked AE networks, we use a chain of fully-connected convolutional layers as the stacked encoders. Image noise and artifacts are suppressed from low-level to high-level step by step in order to preserve essential information in the extracted patches. Moreover, since the pooling layer (down-sampling) after a convolutional layer may discard important structural details, it is abandoned in our encoder. As a result, there are only two types of layers in our encoder: convolutional layers and ReLU units, and the stacked encoders $C_e^i(\mathbf{x}_i)$ can be formulated as
$$C_e^i(\mathbf{x}_i) = \text{ReLU}(\mathbf{W}_i * \mathbf{x}_i + \mathbf{b}_i) \quad i=0,1...,N, \quad (3)$$
where $N$ is the number of convolutional layers, $\mathbf{W}_i$ and $\mathbf{b}_i$ denote the weights and biases respectively, $*$ represents the convolution operator, $\mathbf{x}_0$ is the extracted patch from the input images, and $\mathbf{x}_i (i>0)$ is the extracted features from the previous layers. $\text{ReLU}(x) = \max(0,x)$ is the activation function. After the stacked encoders, the image patches are transformed into a feature space, and the output is a feature vector $\mathbf{x}_N$ whose size is $l_N$.

*3) Stacked decoders (Structural detail recovery)*

Although the pooling operation is removed, a serial of convolutions, which essentially act as noise filters, will still diminish the details of input signals. Inspired by the recent results on semantic segmentation [45, 46, 47] and biomedical image segmentation [48, 49], deconvolutional layers are integrated into our model for recovery of structural details, which can be seen as image reconstruction from extracted features. We use a chain of fully-connected deconvolutional layers to form the stacked decoders for image reconstruction. Since the encoders and decoders should appear in pair, the convolutional and deconvolutional layers are symmetric in the proposed network. To ensure the input and output of the network match exactly, the convolutional and deconvolutional layers must have the same kernel size. Note that the data flow through the convolutional and deconvolutional layers in our framework follows the rule of "FILO" (First In Last Out). As demonstrated in Fig. 1, the first convolution layer corresponds to the last deconvolutional layer, the last convolution layer corresponds to the first deconvolutional layer, and so on. In other words, this architecture is featured by the symmetry of paired convolution and deconvolution layers.

There are two types of layers in our decoder network: deconvolution and ReLU. Thus, the stacked decoders $D_d^i(\mathbf{y}_i)$ can be formulated as:
$$D_d^i(\mathbf{y}_i) = \text{ReLU}(\mathbf{W}_i^{'} \otimes \mathbf{y}_i + \mathbf{b}_i^{'}) \quad i=0,1...,N, \quad (4)$$
where $N$ is the number of deconvolutional layers, $\mathbf{W}_i^{'}$ and $\mathbf{b}_i^{'}$ denote the weights and biases respectively, $\otimes$ represents the deconvolutional operator, $\mathbf{y}_N = x$ is the output feature vector after stacked encoding, $\mathbf{y}_i (N>i>0)$ is the reconstructed feature vector from the previous deconvolutional layer, and $\mathbf{y}_0$ is the reconstructed patch. After stacked decoding, image patches are reconstructed from features, and can be assembled to reconstruct a denoised image.

*4) Residual compensation*

Like the prior art methods [24, 25], convolution will eliminate some image details. Although the deconvolutional layers can recover some of the details, when the network goes deeper this inverse problem becomes more ill-posed, and the accumulated loss could be quite unsatisfactory for image reconstruction. In addition, when the network depth increases the gradient diffusion could make the network difficult to train.

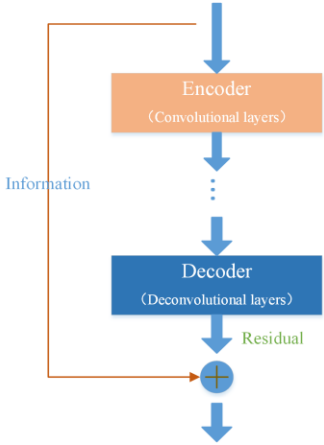

Fig.2. Shortcut in the residual compensation structure.

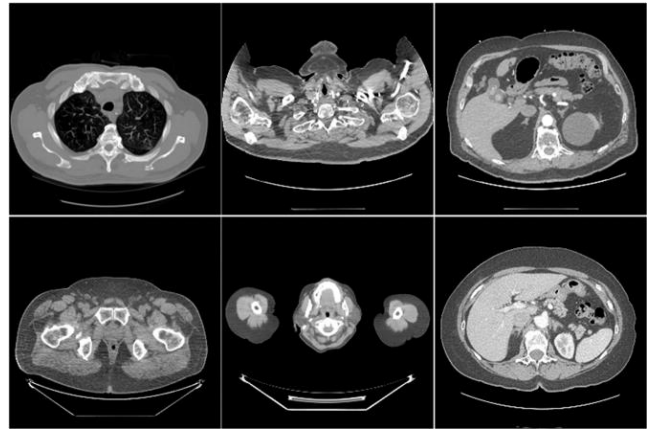

Fig.3. Examples from the normal-dose CT image dataset.

To address the above two issues, similar to deep residual learning [41, 42] we introduce a residual compensation mechanism into the proposed network. Instead of mapping the input to the output solely by the stacked layers, we adopt a residual mapping, as shown in Fig. 2. Defining the input as $I$ and the output as $O$, the residual mapping can be denoted as $F(I) = O - I$, and we use stacked layers to fit this mapping. Once the residual mapping is built, we can reconstruct the original mapping as $R(I) = O = F(I) + I$. Consequently, we transform the direct mapping problem to a residual mapping problem.

There are two benefits associated with the residual mapping. First, it is easier to optimize the residual mapping than optimizing the direct mapping. In other words, it helps avoid the gradient vanishing during training when the network is deep. For example, it would be much easier to train an identity mapping network by pushing the residual to zero than fitting an identity mapping directly. Second, since only the residual is processed by the convolutional and deconvolutional layers, more structural and contrast details can be preserved in the output of the deconvolutional layers, which can significantly enhance the LDCT imaging performance.

The use of shortcut connections in [41, 42] was to solve the difficulty in training so that the shortcut connections were only applied across convolutional layers of the same size. In our work, shortcut connections were used for both preservation of structural details and facilitation of training deeper networks. Furthermore, the symmetric structure of convolution and deconvolution layer pairs was also utilized to keep more details while suppressing image noise and artifacts. The CNN layers in [41] are essentially feedforward long short-term memories (LSTMs) without gates, while our RED-CNN network is in general not composed of the standard feedforward LSTMs.

In [47] and its variants [48, 49], both shortcut connection and deconvolution were used for segmentation. High resolution features were combined with an up-sampled output to improve the image classification. Besides shortcut connection and deconvolution, there are the following new features of the proposed RED-CNN over the networks in [47-49]:

(i). The idea of the autoencoder, which was originally designed for training with noisy samples, was introduced into our model, and convolution and deconvolution layers appeared in pairs;

(ii). To avoid losing details, pooling layer was discarded;

(iii). Convolution layers can be seen as noise filters in our application, but filtering leads to loss in details. Deconvolution and shortcutting in our model were used for detail preservation, and in the experiment section we will separately analyze the improvements due to each of these components. Furthermore, the strides of convolution and deconvolution layers in our model were fixed to 1 to avoid down-sampling.

*5) Training*

The proposed network is an end-to-end mapping from low-dose CT images to normal-dose images. Once the network is configured, the set of parameters, $\Theta = \{\mathbf{W}_i, \mathbf{b}_i, \mathbf{W}_i', \mathbf{b}_i'\}$ of the convolutional and deconvolutional layers should be estimated to build the mapping function $M$. The estimation can be achieved by minimizing the loss $F(D;\Theta)$ between the estimated CT images and the reference NDCT images $X$. Given a set of paired patches $P = \{(X_1, Y_1), (X_2, Y_2), \ldots, (X_K, Y_K)\}$ where $\{X_i\}$ and $\{Y_i\}$ denote NDCT and LDCT image patches respectively, and $K$ is the total number of training samples. The mean squared error (MSE) is utilized as the loss function:

$$F(D;\Theta) = \frac{1}{N}\sum_{i=1}^{N}\|X_i - M(Y_i)\|^2. \quad (5)$$

In this study, the loss function was optimized by Adam [50].

## III. EXPERIMENTAL DESIGN AND RESULTS

*A. Data Sources*

*1) Simulated data*

The normal dose dataset included 7,015 normal-dose CT images of 256×256 pixels per image from 165 patients downloaded from the National Biomedical Imaging Archive (NBIA). Different parts of the human body were included for diversity. Some typical images are in Fig. 3. The corresponding LDCT images were produced by adding Poisson noise into the sinograms simulated from the normal-dose images. With the assumed use of a monochromatic source, the projection measurements from a CT scan follow the Poisson distribution, which can be expressed as

$$z_i \sim \text{Poisson}\left\{b_i e^{-l_i} + r_i\right\}, i = 1,...,I \qquad (6)$$

where $z_i$ is the measurement along the $i$-th ray path. $l_i$ is the line integral of attenuation coefficients, $b_i$ is the blank scan factor, and $r_i$ stands for read-out noise. For the simulation, the noise level can be controlled by the blank scan factor $b_i$. In our initial studies, $b_i$ was uniformly set to $10^5$ photons and denoted as $b_0 = b_i = 10^5, i = 1,...,I$. Siddon's ray-driven method [51] was used to generate the projection data in fan-beam geometry. The source-to-rotation center distance was 40 cm while the detector-to-rotation center was 40 cm. The image region was set to 20 cm × 20 cm. The detector width was 41.3 cm containing 512 detector elements. There were 1,024 viewing angles uniformly distributed over a full scan range.

Since direct processing of an entire CT image is intractable, RED-CNN was applied to image patches. Our method benefits from the patch extraction since patch-based processing represents the local details required for optimal denoising and the number of samples were greatly increased for the training purpose. Deep learning methods require a large amount of training samples, but collecting medical images is usually limited by the complicated formalities addressing multiple factors such as the patient's privacy.

In the training step, 200 normal-dose and corresponding simulated low-dose images were randomly selected as the training set. Also, 100 image pairs were randomly selected as the testing set. Images from the patients in the training set were not in the testing set.

*2) Clinical data*

To validate the clinical performance of RED-CNN, a real clinical database was used, which was authorized by Mayo Clinics for "*the 2016 NIH-AAPM-Mayo Clinic Low Dose CT Grand Challenge*". The dataset contained 2,378 3mm thickness full and quarter dose 512×512 CT images from 10 patients [37]. The network was trained with a subset of full dose and quarter dose image pairs. The rest of the image pairs were respectively used as the testing set and the gold standard. For fairness, cross-validation was utilized in the testing phase. While testing on CT images from each patient, the images from the other 9 patients were involved in the training phase.

There are three reasons why we used both simulated and clinical data. First, the database from NBIA is more diverse than that at Mayo. It includes more body parts than the database at Mayo, and therefore more realistically reflects clinical imaging applications. Second, different from the clinical dataset, which has both full-dose datasets and their corresponding quarter-dose counterparts, the low-dose images from NBIA were simulated by adding Poisson noise into the simulated sinograms. By doing so, we can control the noise levels of the data to simulate different doses, and we can evaluate the robustness of our method as described in III D (5). Third, the experiments on simulated data were focused on the improvement of image quality with our model while the experiments with clinical data targeted clinical tasks, such as low contrast lesion detection.

*B. Parameter selection*

The patch size $\eta$ was set to 55×55 with the sliding interval of 4 pixels. After the patch extraction, the number of training patch pairs reached $10^6$. Furthermore, three kinds of transformation operations, including rotation (by 45 degrees), flipping (vertical and horizontal) and scaling (scale factors were 2 and 0.5), were used for data augmentation. The network was implemented in Caffe. In our experiments, we evaluated several parameter combinations and finalized the parameter settings as follows. The base learning rate was set to $10^{-4}$, and slowly decreased down to $10^{-5}$. The convolution and deconvolution kernels were initialized with random Gaussian distributions with zero mean and standard deviation 0.01. The filter number of last layer was set to 1 and the others were set to 96. The kernel size of all layers was set to 5×5. The strides of convolution and deconvolution were set to 1 with no padding. All the experiments were performed with MATLAB 2015b on a PC (Intel i7 6700K CPU and 16 GB RAM). The training stage is time-consuming for traditional CPU implementation. A common way for acceleration is to work in a parallel manner. In our work, the training of RED-CNN was performed on a graphic processing unit card (GTX 1080). Although the training was done on patches, the proposed network can process images of arbitrary sizes. All the testing images were simply fed into the network, without decomposition.

The three metrics, including the root mean square error (RMSE), peak signal to noise ratio (PSNR) and structural similarity index measure (SSIM), were chosen for quantitative assessment of image quality.

Five different state-of-the-art methods were compared against our RED-CNN, including TV-POCS [6], K-SVD [18], BM3D [20], CNN10 [37], and KAIST-Net [38]. Dictionary learning and BM3D are two most popular image-based denoising methods already applied for LDCT. ASD-POCS is a widely used iterative reconstruction method under the TV regularization. CNN10 is a simplified version of the proposed RED-CNN without shortcuts and deconvolutional layers. It also can be viewed as a deeper version of the CNN-based LDCT restoration model [37]. KAIST-Net is the most recently proposed CNN-based LDCT denoising method. It can be considered as a deepened variant of the lightweight CNN model [37]. The parameters of these competing methods were set per the suggestions from the original papers.

*C. Experimental Results*

*1) Simulated data*

Two representative slices from the testing dataset were used to demonstrate the performance of RED-CNN, which are through the chest and abdominal regions respectively. It can be seen that the normal-dose images from different scan protocols contained different noise levels. Fig. 4 shows our results from the chest image. In Fig. 4(b), there was high image noise and streaking artifacts adjacent to structures with high attenuation coefficients such as bones. All applied methods suppressed image noise to various degrees. However, in Fig. 4(c), TV-POCS suffered from a blocky effect and also smoothened some important small structures in the lungs. K-SVD and BM3D preserved more details than TV-POCS, but there were artifacts near the bones. CNN10, KAIST-Net and RED-CNN eliminated most image noise and artifacts while preserving the structural features better than the other methods. Furthermore, RED-CNN discriminated low contrast regions in the best way. Fig. 5 shows the zoomed images over a region of interest (ROI). Clearly, the blood vessels in the lungs, highlighted by the blue arrow, were smoothened by TV-POCS in Fig. 5(c). The other methods could identify these details to different extents, and the details were retained without blurring in Fig. 5(h). Meanwhile, in Fig. 5(d) and (e) the streaking artifacts were evident near the bone, marked by the red arrow. To further show the merits of RED-CNN, the absolute difference images relative to the original image are in Fig. 6. It can be clearly observed that RED-CNN yielded the smallest difference from the original normal-dose image, preserving all details and suppressing most noise and artifacts.

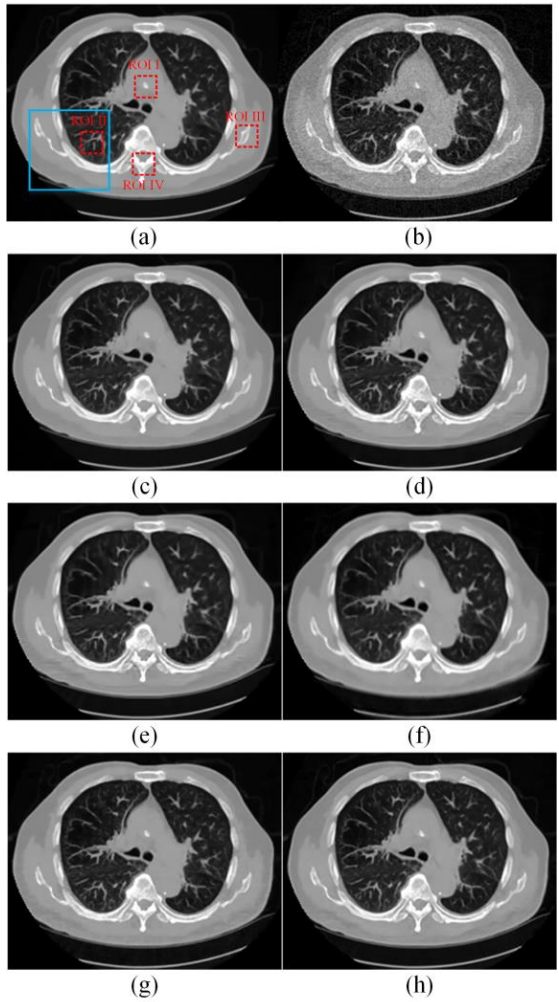

Fig. 4. Results from the chest image for comparison. (a) NDCT, (b) LDCT, (c) TV-POCS, (d) K-SVD, (e) BM3D, (f) CNN10, (g) KAIST-Net, and (h) RED-CNN. The blue box indicates the region zoomed in Fig. 5. The red dotted boxes define several ROIs.

For quantitative evaluation, four ROIs were chosen as highlighted by red dotted boxes in Fig. 4(a). The results are in Fig. 7. The quantitative results followed similar trends per visual inspection. The RED-CNN had the lowest RMSE and the highest PSNR/SSIM for all the ROIs.

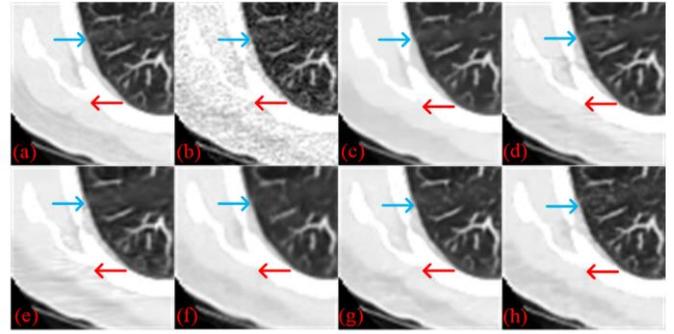

Fig. 5. Zoomed parts over the region of interest (ROI) marked by the blue box in Fig. 4(a). (a) NDCT, (b) LDCT, (c) TV-POCS, (d) K-SVD, (e) BM3D, (f) CNN10, (g) KAIST-Net, and (h) RED-CNN ((a)-(h) from Fig. 4(a)-(h)). The arrows indicate two regions for visual differences.

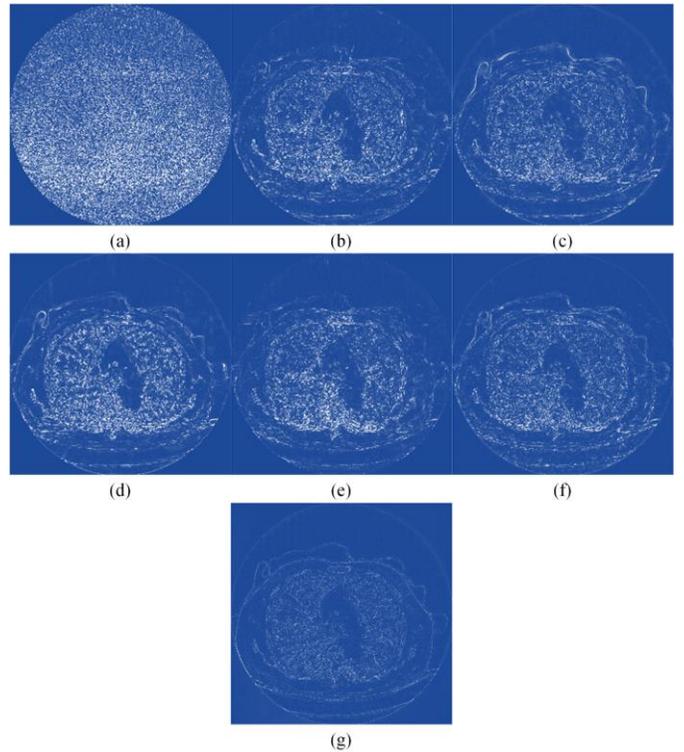

Fig. 6. Absolute difference images relative to the NDCT image. (a) LDCT, (b) TV-POCS, (c) K-SVD, (d) BM3D, (e) CNN10, (f) KAIST-Net, and (g) RED-CNN.

Fig. 8 presents the results from the abdominal image. Since the image quality of the original normal-dose image (Fig. 8(a)) is worse than the chest image (Fig. 4(a)), the simulated LDCT image suffered from severe deterioration and many structures cannot be distinguished in Fig. 8(b). TV-POCS and K-SVD cannot recover the image well. The blocky effect appeared in Fig. 8(c). BM3D eliminated most noise but the artifacts close to the spine were evident. CNN10, KAIST-Net and RED-CNN suppressed most of the noise and artifacts but the result in Fig. 9(f) and (g) suffered a bit from over-smoothing, which is consistent to the previous results with a lightweight CNN as reported in [37]. The red arrows indicate several noticeable structural differences between different methods. The linear high attenuation structure in the liver likely representing a contrast enhanced blood vessel was best retained by RED-CNN in Fig. 9. The low attenuation pseudo lesions noted in the posterior aspect of the liver on other techniques were not seen on RED-CNN. The thin and subtle right adrenal gland was best appreciated on the RED-CNN image as well. Finally, margins of different tissues were also better delineated and retained on the RED-CNN image.

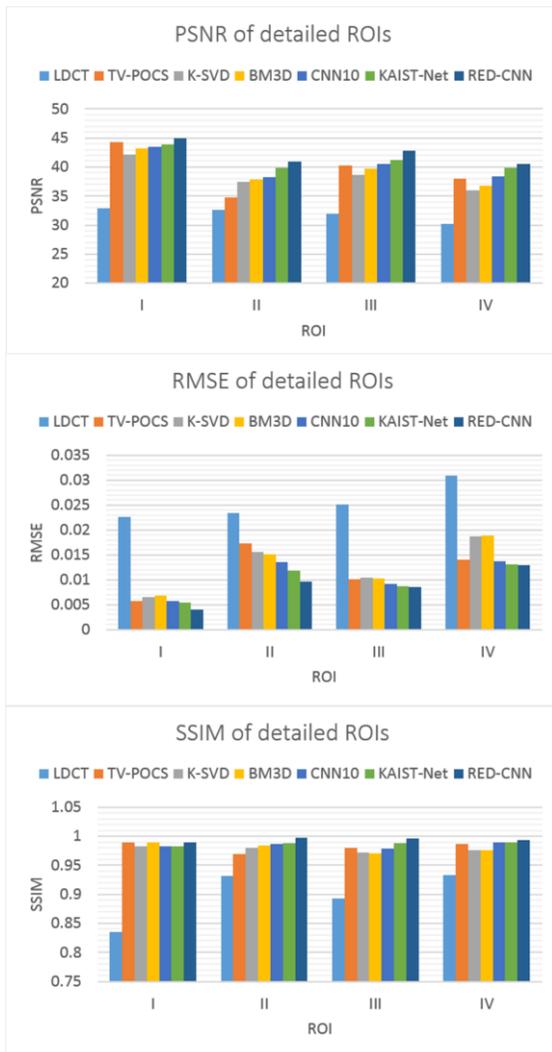

Fig. 7. Performance comparison of the six algorithms over the ROIs marked in Fig. 4(a) in terms of the selected metrics.

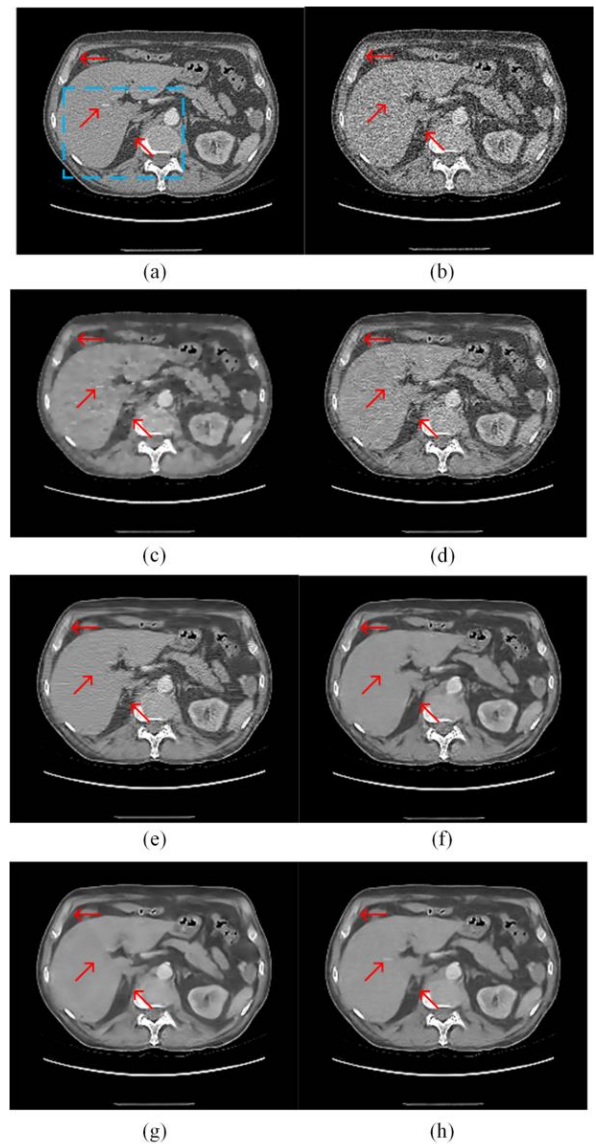

Fig. 8. Results from the abdominal image for comparison. (a) NDCT, (b) LDCT, (c) TV-POCS, (d) K-SVD, (e) BM3D, (f) CNN10, (g) KAIST-Net, and (h) RED-CNN. The arrows indicate three regions to observe the visual effects.

The quantitative results for the whole images restored using these methods are listed in Table I. It can be seen that RED-CNN obtained the best scores on all the indexes.

Table II gives the mean measurements for all the 100 images from the testing dataset. Again, it can be seen that the proposed RED-CNN outperformed the state-of-the-art methods in terms of each of the metrics.

*2) Clinical data*

Two representative slices from real clinical CT scans were chosen to evaluate the performance of the proposed RED-CNN. The results are in Figs. 10-13. In Figs. 10 and 12, it is clear that RED-CNN delivered the best performance in terms of both noise suppression and structure preservation. In the zoomed parts in Figs. 11 and 13, the low-contrast liver lesions highlighted by the red circles were processed using our and others' methods, and our proposed method gave the best image quality. TV-POCS and K-SVD generated some artifacts and lowered the detectability of lesions. BM3D blurred the low-contrast lesions. All the methods except KAIST-Net and RED-CNN could not distinguish the contrast enhanced blood vessel marked by the red arrow in Fig. 11. Although KAIST-Net had a performance similar to that of our method, KAIST-Net smoothened the low contrast lesions in Fig. 11(g) while our RED-CNN preserved the edges very well in Fig. 11(h). Meanwhile, two tiny focal low attenuation lesions were hard to detect in Fig. 12(f) and (g) but they can be noticed in Fig. 12(h).

Table III summarizes the quantitative results from the aforementioned two images. RED-CNN gave better performance in terms of most of the metrics than the other methods. Table IV shows the quantitative results on the full cross validation in terms of means ± SDs (average scores ± standard deviations). All our visual observations are supported by the quantitative evaluation as shown in Table IV.

For qualitative evaluation, 10 LDCT images with lesions and the processed images using different methods were selected for a reader study. Artifact reduction, noise suppression, contrast retention, lesion discrimination, and overall quality were used as subjective indicators on the five-point scale (1 = unacceptable and 5 = excellent). Two radiologists (R1 and R2) with 6 and 8 years of clinical experience respectively evaluated these images independently to provide their scores. The NDCT images were used as the gold standard. For each set of images, the scores were reported as means ± SDs (average scores of the two radiologists ±standard deviations). The student $t$-test with $p<0.05$ was performed. The statistical results are in Table V. For all the five indicators, the LDCT images had the lowest scores due to their severe image quality degradation. All the LDCT methods significantly improved the scores. KAIST-Net and RED-CNN produced substantially higher scores than the other methods, and RED-CNN performed slightly better than KAIST-Net and ran significantly faster than KAIST-Net in both the training and testing processes.

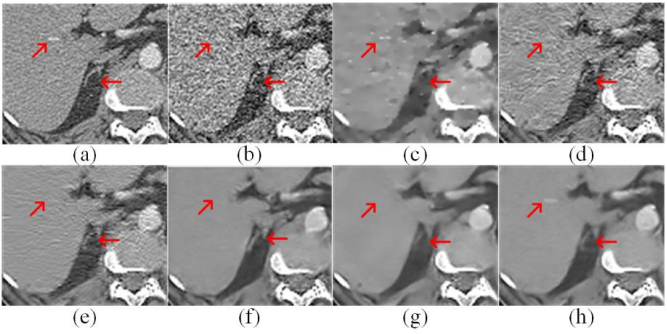

Fig. 9. Zoomed ROI images from Fig. 8. (a) NDCT, (b) LDCT, (c) TV-POCS, (d) K-SVD, (e) BM3D, (f) CNN10, (g) KAIST-Net, and (h) RED-CNN ((a)-(h) from Fig. 8(a)-(h)). The arrows indicate two regions containing features revealed differently by the competing algorithms.

TABLE I
QUANTITATIVE RESULTS ASSOCIATED WITH DIFFERENT ALGORITHMS FOR THE ABDOMINIAL IMAGE.

|  | PNSR | RMSE | SSIM |
|---|---|---|---|
| LDCT | 34.3094 | 0.0193 | 0.8276 |
| TV-POCS | 37.5485 | 0.0133 | 0.8825 |
| K-SVD | 38.3841 | 0.0120 | 0.9226 |
| BM3D | 38.9903 | 0.0112 | 0.9295 |
| CNN10 | 38.9907 | 0.0104 | 0.9288 |
| KAIST-Net | 38.9908 | 0.0102 | 0.9283 |
| RED-CNN | **39.1959** | **0.0097** | **0.9339** |

TABLE II
QUANTITATIVE RESULTS (MEAN±SDs) ASSOCIATED WITH DIFFERENT ALGORITHMS FOR THE IMAGES IN THE TESTING DATASET

|  | PNSR | RMSE | SSIM |
|---|---|---|---|
| LDCT | 36.3975±5.24 | 0.0158±0.0062 | 0.8644±0.0754 |
| TV-POCS | 41.5021±2.11 | 0.0087±0.0010 | 0.9498±0.0126 |
| K-SVD | 40.8445±2.54 | 0.0096±0.0013 | 0.9447±0.0168 |
| BM3D | 41.5358±2.09 | 0.0088±0.0010 | 0.9509±0.0127 |
| CNN10 | 41.9892±2.10 | 0.0082±0.0011 | 0.9658±0.0129 |
| KAIST-Net | 42.2746±2.05 | 0.0078±0.0009 | 0.9688±0.0098 |
| RED-CNN | **43.7871±2.01** | **0.0069±0.0007** | **0.9754±0.0086** |

### D. Model and Performance Trade-Offs

In this subsection, several critical factors of the proposed RED-CNN were examined, including deconvolutional decoder, shortcut connection, number of the layers, patch size and robustness with respect to the training and testing datasets. Computational costs were also discussed. The data used to plot the curves in the following sections were randomly selected from the NBIA dataset (average values from 40 images).

*1) Deconvolutional Decoder*

Different from the traditional convolutional layers, deconvolutional layers, also referred to as learnable up-sampling layers, can produce multiple outputs with a single input. This architecture has been successful in sematic segmentation coupled with convolutional layers [45-49]. With traditional fully-connected CNNs [27, 28, 37], some important details could be lost in the convolution. That is why the number of layers of these CNNs are usually less than 5, for low-level tasks, such as denoising, deblurring and super resolution [24-30]. In our proposed model, we have balanced the conventional CNN layers with an equal number of deconvolutional layers, forming the network capable of bringing the details back to the image of the original size. In our network, pooling and unpooling operations are avoided to keep structural information in the images. We assessed the performance of the networks with and without deconvolutional layers as shown in Fig. 14. It can be seen that our model with the deconvolution mechanism performed better than the fully-convolutional counterpart.

*2) Shortcut Connection*

Shortcut connection is another trick we used to improve the structural preservation. It has been proven useful in the both high- and low-level tasks, such as image recognition [40, 41, 47-49] and image restoration [31]. We evaluated the impact of shortcuts in the proposed RED-CNN. The results are in Fig. 15. The model with shortcuts produced better PSNR and RMSE values and converged more rapidly.

*3) Number of Layers*

Recent studies suggested that deeper network architectures, especially CNN-based models, produced better performance for image recognition [41, 42]. Here we investigated the trade-off between performance and the number of layers by testing the use of 10, 20 and 30 layers. The quantitative results are in Fig. 16. It can be seen that the differences were not easily noticeable. This observation is consistent to the statement in [28, 31] that deeper networks do not always result in better performance in low-level image processing tasks. Although the utilization of shortcut connections enables much more layers than the preliminary explorations [28, 37], it seems that the enhanced performance by adding more layers is limited, and better understanding for training dynamics of deep networks may help overcome this bottleneck.

*4) Patch Size*

For CNN-based image restoration [24, 28, 31], in the training stage image patch pairs were used, and in the testing stage the whole images were directly fed into the trained network. Training with patches can enhance the detectability of perceptual differences of local regions, and the amount of samples are significantly boosted. Once the filters in each layer are trained well, due to the property of convolution operators, there is no difference between different patch sizes with which the network is fed. Here we tried to sense the impact with different training patch sizes. The result is in Fig. 17. We increased the patch size from 55 to 100 and there was no significant difference shown up. Based on this observation, in our experiments we fixed the patch size to 55×55 for better trade-off between training time and imaging performance.

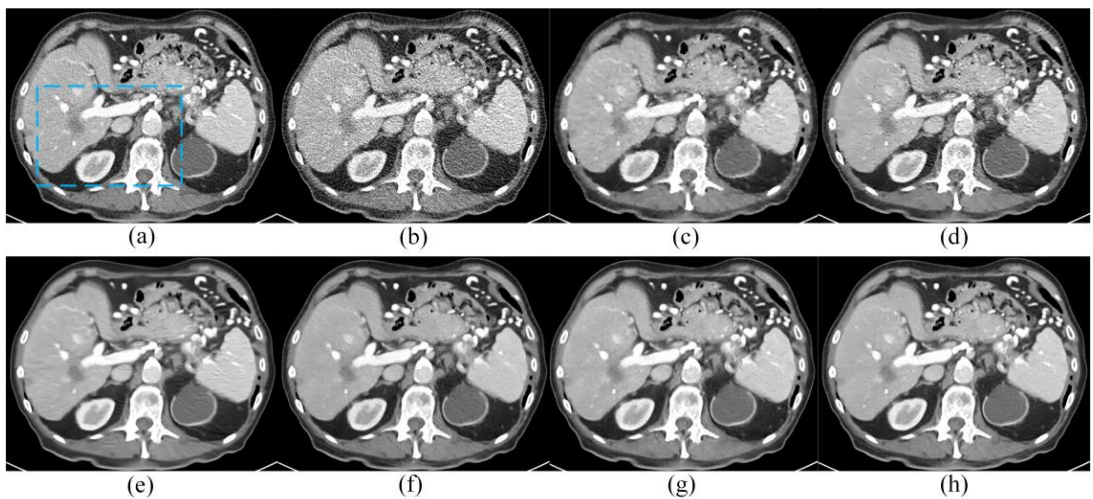

Fig. 10. Results from the abdominal image with a metastasis in the liver for comparison. (a) NDCT, (b) LDCT, (c) TV-POCS; (d) K-SVD, (e) BM3D, (f) CNN10, (g) KAIST-Net, and (h) RED-CNN.

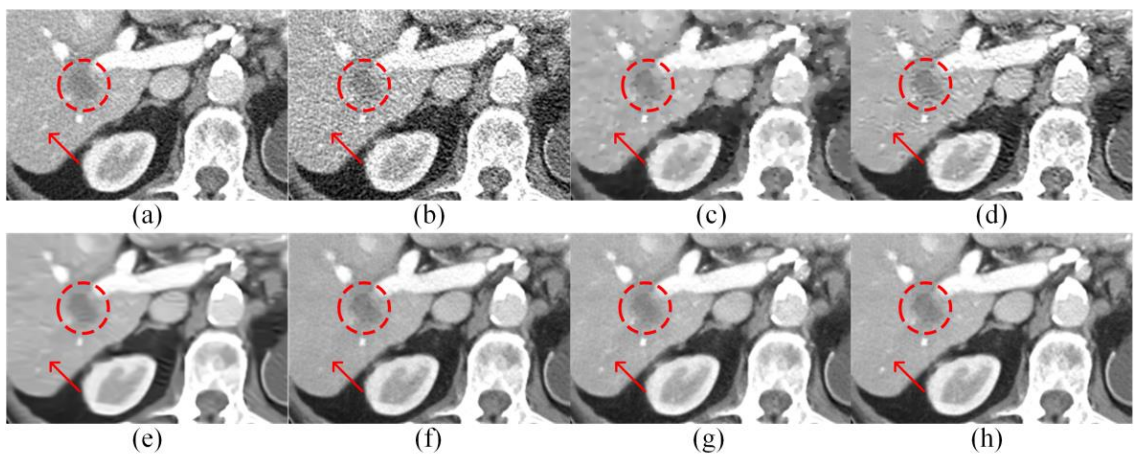

Fig. 11. Zoomed parts from Fig. 10. (a) NDCT, (b) LDCT, (c) TV-POCS, (d) K-SVD, (e) BM3D, (f) CNN10, (g) KAIST-Net, and (h) RED-CNN. The circle indicates the lesion while the arrow points to the contrast enhanced blood vessel.

*5) Performance Robustness*

In our experiments, the noise level was fixed, and the corresponding network was trained under the assumption of a uniform noise level. In practice, it is impossible to train with different parameter sets subject to different noise levels. Since the IR methods usually have several parameters, it is inconvenient to explore the possible parameter space for optimal image quality. We believe that the proposed RED-CNN model is robust for different noise levels. To show the robustness of RED-CNN, several combinations of noise levels in the training and testing datasets were simulated to generate the quantitative results in Table VI. In this table, the training dataset of CNN10, KAIST-Net and RED-CNN were made for $b_0 = 10^5$. CNN10+, KAIST-Net+ and RED-CNN+ denote the same networks as CNN10, KAIST-Net and RED-CNN with randomly mixed training data at different noise levels for $b_0 = 10^5$, $b_0 = 5 \times 10^5$ and $b_0 = 5 \times 10^4$. It is clear that RED-CNN+ obtained the best performance in most of the situations, which means that if an accurate noise level cannot be determined, a good solution is to train the network with mixed data at possible noise levels. In the column 'RED-CNN', it is seen that even if training is done with a single noise level, RED-CNN is still competitive in handling the cases of inconsistent noisy data.

*6) Computational Cost*

The computational cost is another advantage of deep learning. Although the training is time-consuming, it can be improved with GPU. For the dataset involved in our experiments, training took about 4 hours for about $10^6$ patches and 12 hours for about $10^7$ patches. CNN10 has the same number of layers as that of RED-CNN, but without the shortcut, it took more time in training. It ran 6 hours for about $10^6$ patches and 15 hours for about $10^7$ patches. KAIST-Net has a complex architecture with 26 layers and 15 channel inputs, and as a result the training time was much longer (also implemented in Caffe). For $10^6$ and $10^7$ patches, it took 12 hours and 30 hours respectively. The other methods, especially for iterative reconstruction, do not need a training process, but the execution time is much longer than CNN10, KAIST-Net and RED-CNN. In this study, the average execution times for ASD-POCS, K-SVD, BM3D, CNN10, KAIST-Net and RED-CNN are 21.36, 38.45, 4.22, 3.22, 30.22 and 3.68 seconds respectively. Actually, after the network is trained offline, the proposed model is much more efficient than any other methods in terms of execution time.

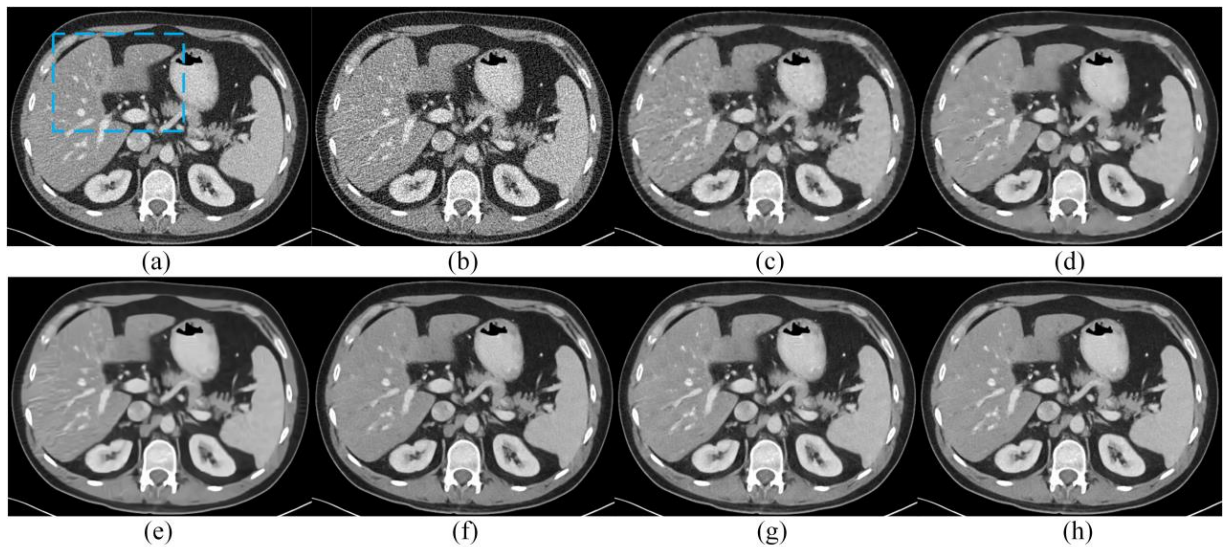

Fig. 12. Results from the abdominal image with two focal fatty sparings in the liver for comparison. (a) NDCT, (b) LDCT, (c) TV-POCS, (d) K-SVD, (e) BM3D, (f) CNN10, (g) KAIST-Net, and (h) RED-CNN.

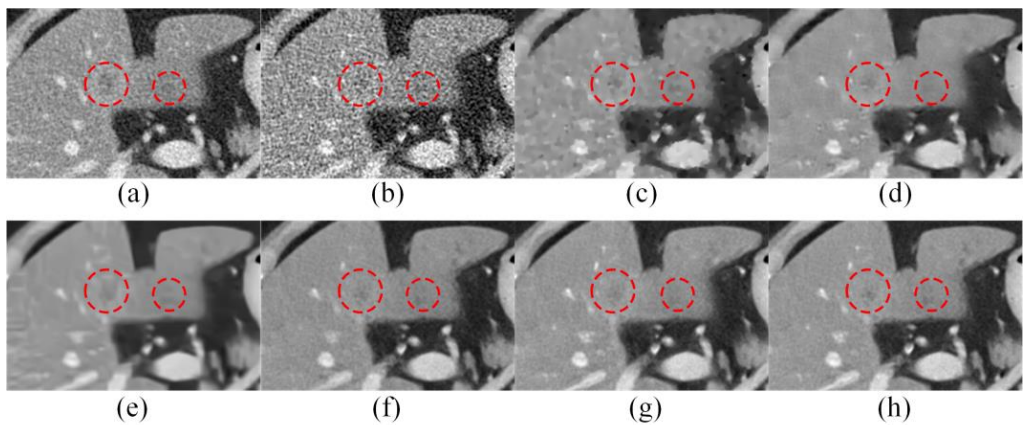

Fig. 13. Zoomed parts from Fig. 12. (a) NDCT, (b) LDCT, (c) TV-POCS, (d) K-SVD, (e) BM3D, (f) CNN10, (g) KAIST-Net, and (h) RED-CNN.

TABLE III
QUANTITATIVE RESULTS ASSOCIATED WITH DIFFERENT ALGORITHMS FOR FIGS. 10 AND 12.

|  | Fig. 10 | | | Fig. 12 | | |
| --- | --- | --- | --- | --- | --- | --- |
|  | PNSR | RMSE | SSIM | PNSR | RMSE | SSIM |
| LDCT | 40.2252 | 0.0097 | 0.9292 | 39.4243 | 0.0107 | 0.9145 |
| TV-POCS | 42.582 | 0.0074 | 0.9649 | 42.2391 | 0.0077 | 0.9624 |
| K-SVD | 43.8516 | 0.0064 | 0.9678 | 43.6871 | 0.0065 | 0.9668 |
| BM3D | 44.0152 | 0.0063 | 0.9696 | 43.7136 | 0.0065 | 0.9672 |
| CNN10 | 45.1678 | 0.0057 | 0.9712 | 44.2286 | 0.0061 | 0.9698 |
| KAIST-Net | 45.1845 | **0.0053** | 0.9735 | 44.3247 | **0.0058** | 0.9708 |
| RED-CNN | **45.2616** | 0.0055 | **0.9764** | **44.6466** | 0.0059 | **0.9718** |

TABLE IV
QUANTITATIVE RESULTS (MEAN±SDs) ASSOCIATED WITH DIFFERENT ALGORITHMS ON THE FULL CROSS VALIDATION

|  | PNSR | RMSE | SSIM |
| --- | --- | --- | --- |
| LDCT | 39.4314±1.5206 | 0.0109±0.0021 | 0.9122±0.0280 |
| TV-POCS | 41.7496±1.1522 | 0.0083±0.0012 | 0.9535±0.0143 |
| K-SVD | 42.7203±1.4260 | 0.0074±0.0014 | 0.9531±0.0167 |
| BM3D | 42.7661±1.0471 | 0.0073±0.0009 | 0.9563±0.0125 |
| CNN10 | 43.6561±1.1323 | 0.0066±0.0009 | 0.9664±0.0100 |
| KAIST-Net | 43.9668±1.2169 | 0.0064±0.0009 | 0.9688±0.0110 |
| RED-CNN | **44.4187±1.2118** | **0.0060±0.0009** | **0.9705±0.0087** |

## IV. CONCLUSION

In brief, we have designed a symmetrical convolutional and deconvolutional neural network, aided by shortcut connections. Two well-known databases have been utilized to evaluate and validate the performance of our proposed RED-CNN in comparison with the state of the art methods. The simulated and clinical results have demonstrated a great potential of deep learning for noise suppression, structural preservation, and lesion detection at a high computational speed. In the future, we plan to optimize RED-CNN, extend it to higher dimensional cases such as 3D reconstruction, dynamic/spectral CT reconstruction, and adapt the ideas to other imaging tasks or even other imaging modalities.

TABLE V
STATISTICAL ANALYSIS OF SUBJECTIVE QUALITY SCORES FOR DIFFERENT ALGORITHMS (MEAN ± SDs).

|  | NDCT | LDCT | TV-POCS | K-SVD | BM3D | CNN10 | KAIST-Net | RED-CNN |
|---|---|---|---|---|---|---|---|---|
| Artifact reduction | 3.55±0.26 | 1.88±0.33* | 2.47±0.28* | 2.74±0.31* | 2.94±0.28* | 3.39±0.24 | 3.45±0.24 | **3.50±0.23** |
| Noise suppression | 3.67±0.28 | 2.04±0.31* | 2.59±0.32* | 2.90±0.35* | 3.11±0.29* | 3.55±0.28 | 3.57±0.35 | **3.64±0.21** |
| Contrast retention | 3.22±0.22 | 1.95±0.34* | 2.27±0.29* | 2.77±0.35* | 2.84±0.24* | 3.01±0.32 | 3.08±0.24 | **3.17±0.28** |
| Lesion discrimination | 3.24±0.23 | 1.75±0.29* | 2.24±0.30* | 2.67±0.32* | 2.70±0.33* | 3.15±0.24 | 3.12±0.28 | **3.21±0.28** |
| Overall quality | 3.39±0.24 | 1.91±0.32* | 2.33±0.34* | 2.64±0.29* | 2.72±0.29* | 2.99±0.28 | 3.08±0.26 | **3.25±0.21** |

\* indicates P < 0.05, which means significantly different.

TABLE VI
QUANTITATIVE RESULTS (MEAN) ASSOCIATED WITH DIFFERENT ALGORITHMS FOR VARIOUS COMBINATIONS OF NOISE LEVELS.

| Noise level of testing data | | TV-POCS | K-SVD | BM3D | CNN10 | CNN10+ | KAIST-Net | KAIST-Net+ | RED-CNN | RED-CNN+ |
|---|---|---|---|---|---|---|---|---|---|---|
| $b_0 = 5\times10^5$ | PSNR | 44.8030 | 44.0567 | 44.2798 | 44.4584 | 44.4755 | 44.5542 | 44.9645 | 44.8947 | **45.1584** |
|  | RMSE | 0.0061 | 0.0064 | 0.0063 | 0.0062 | 0.0061 | 0.0061 | 0.0058 | 0.0058 | **0.0056** |
|  | SSIM | 0.9735 | 0.9778 | 0.9796 | 0.9792 | 0.9791 | 0.9794 | 0.9827 | 0.9812 | **0.9825** |
| $b_0 = 10^5$ | PSNR | 41.5021 | 40.8445 | 41.5358 | 41.9892 | 42.1248 | 42.2746 | 43.88781 | 43.7871 | **44.1024** |
|  | RMSE | 0.0087 | 0.0096 | 0.0088 | 0.0082 | 0.0081 | 0.0078 | 0.0075 | 0.0069 | **0.0065** |
|  | SSIM | 0.9498 | 0.9447 | 0.9509 | 0.9658 | 0.9655 | 0.9688 | 0.9765 | 0.9754 | **0.9778** |
| $b_0 = 5\times10^4$ | PSNR | 39.7729 | 38.9090 | 39.8928 | 40.5487 | 40.7954 | 40.9857 | 41.8451 | 41.4278 | **42.9892** |
|  | RMSE | 0.0106 | 0.0121 | 0.0106 | 0.0095 | 0.0094 | 0.0092 | **0.0084** | 0.0089 | 0.0091 |
|  | SSIM | 0.9221 | 0.9296 | 0.9492 | 0.9588 | 0.9592 | 0.9601 | 0.9688 | 0.9654 | **0.9727** |

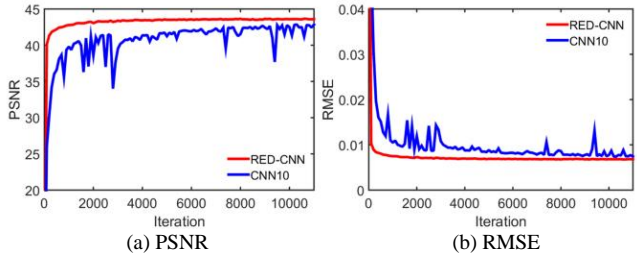

(a) PSNR  (b) RMSE

Fig. 14. PSNR and RMSE values on the testing dataset during training. Our network exhibits a better performance than CNN10. The display ranges of PSNR and RMSE are [20 45] and [0 0.04] respectively.

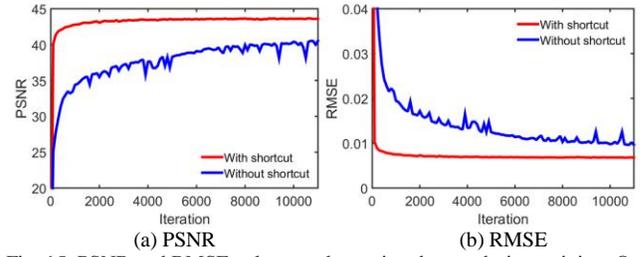

(a) PSNR  (b) RMSE

Fig. 15. PSNR and RMSE values on the testing dataset during training. Our network with shortcut connections exhibits a better performance than the one without shortcuts. The display ranges of PSNR and RMSE are [20 45] and [0 0.04] respectively.

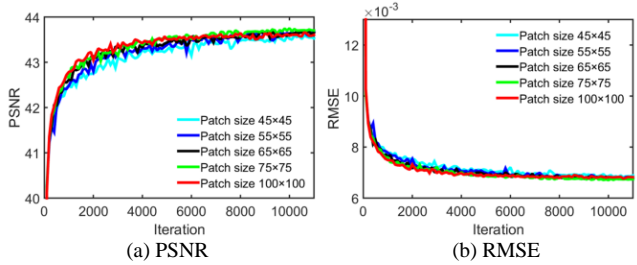

(a) PSNR  (b) RMSE

Fig. 17. PSNR and RMSE values on the testing dataset during training based on different patch sizes. The display ranges of PSNR and RMSE are [40 44] and [0.006 0.013] respectively.

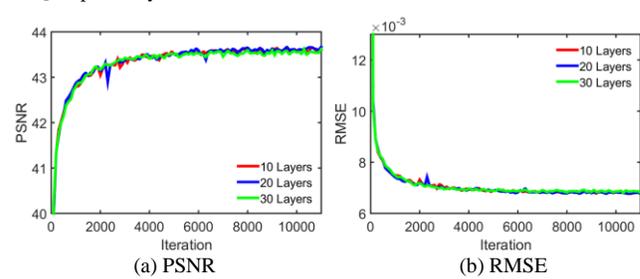

(a) PSNR  (b) RMSE

Fig. 16. PSNR and RMSE values on the testing dataset during training based on different numbers of layers. The display ranges of PSNR and RMSE are [40 44] and [0.006 0.013] respectively.